\documentclass[reprint, aip, cha]{revtex4-1}
\usepackage{amsmath}
\usepackage{upgreek}
\usepackage{graphicx}
\usepackage{lmodern}
\usepackage{mathpazo}
\usepackage[colorlinks]{hyperref}
\hypersetup
{
colorlinks=true,
linkcolor=black,
urlcolor=black,
citecolor=black
}

\def\braket#1{\mathinner{\langle{#1}\rangle}}
\def\ll#1{}
\newcommand{\mean}[3]{\braket{\left|#1(#2,#3)\right|^{2}}}
\newcommand{\intensity}[3]{\left|\psi_{#1}(#2,#3)\right|^{2}}

\begin{document}

\title{\Large Quantum transport simulations in a programmable nanophotonic processor}

\author{Nicholas C. Harris$^\star$}
\affiliation{Department of Electrical Engineering and Computer Science, Massachusetts Institute of Technology, 77 Massachusetts Avenue, Cambridge, MA 02139, USA}
\author{Gregory R. Steinbrecher}
\affiliation{Department of Electrical Engineering and Computer Science, Massachusetts Institute of Technology, 77 Massachusetts Avenue, Cambridge, MA 02139, USA}
\author{Jacob Mower}
\affiliation{Department of Electrical Engineering and Computer Science, Massachusetts Institute of Technology, 77 Massachusetts Avenue, Cambridge, MA 02139, USA}
\author{Yoav Lahini}
\affiliation{Harvard John A. Paulson School of Engineering and Applied Sciences, Harvard University, Cambridge, MA 02138, USA}
\author{Mihika Prabhu}
\affiliation{Department of Electrical Engineering and Computer Science, Massachusetts Institute of Technology, 77 Massachusetts Avenue, Cambridge, MA 02139, USA}
\author{Darius Bunandar}
\affiliation{Department of Physics, Massachusetts Institute of Technology, 77 Massachusetts Avenue, Cambridge, MA 02139, USA}
\author{Changchen Chen}
\affiliation{Department of Electrical Engineering and Computer Science, Massachusetts Institute of Technology, 77 Massachusetts Avenue, Cambridge, MA 02139, USA}
\author{Franco N. C. Wong}
\affiliation{Department of Electrical Engineering and Computer Science, Massachusetts Institute of Technology, 77 Massachusetts Avenue, Cambridge, MA 02139, USA}
\author{Tom Baehr-Jones}
\affiliation{Elenion Technologies, 171 Madison Avenue, Suite 1100, New York, NY 10016, USA}
\author{Michael Hochberg}
\affiliation{Elenion Technologies, 171 Madison Avenue, Suite 1100, New York, NY 10016, USA}
\author{Seth Lloyd}
\affiliation{Department of Mechanical Engineering, Massachusetts Institute of Technology, 77 Massachusetts Avenue, Cambridge, MA 02139, USA}
\author{Dirk Englund}
\affiliation{Department of Electrical Engineering and Computer Science, Massachusetts Institute of Technology, 77 Massachusetts Avenue, Cambridge, MA 02139, USA}
\affiliation{\small{$^\star$\textbf{Corresponding author.}}}

\begin{abstract}
	Environmental noise and disorder play critical roles in quantum particle and wave transport in complex media, including solid-state and biological systems. Recent work has predicted that coupling between noisy environments and disordered systems, in which coherent transport has been arrested due to localization effects, could actually enhance transport. Photonic integrated circuits are promising platforms for studying such effects, with a central goal being the development of large systems providing low-loss, high-fidelity control over all parameters of the transport problem. Here, we fully map the role of disorder in quantum transport using a nanophotonic processor consisting of a mesh of 88 generalized beamsplitters programmable on microsecond timescales. Over 64,400 transport experiments, we observe several distinct transport regimes, including environment-assisted quantum transport and the ``quantum Goldilocks'' regime in strong, statically disordered discrete-time systems. Low loss and high-fidelity programmable transformations make this nanophotonic processor a promising platform for many-boson quantum simulation experiments.
\end{abstract}

\maketitle

{
	\begin{figure*}[t]
	\includegraphics[width=7.2in]{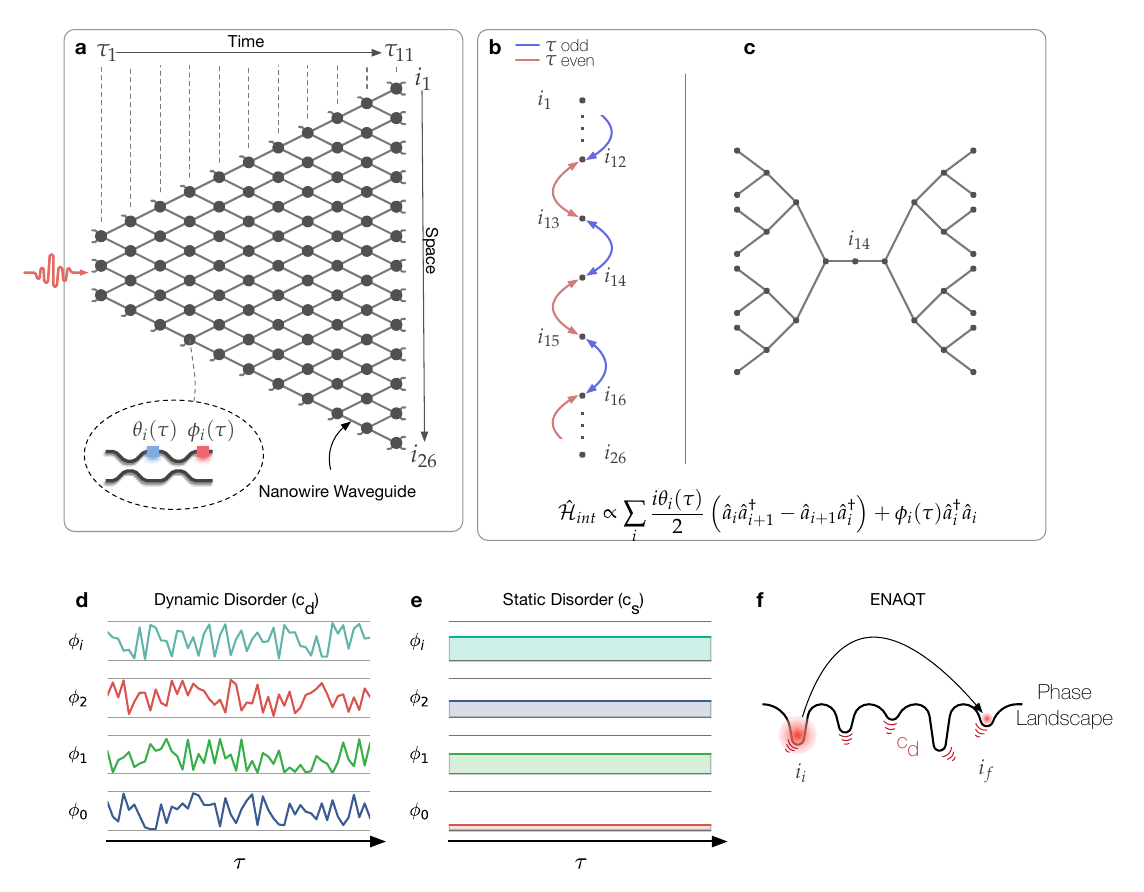}
	\centering
	\caption
	{
		\textsf
		{
		\textbf{A quantum transport simulator.}
		(a) Schematic of programmable nanophotonic processor. Dark lines are silicon nanowire waveguides; circles are programmable Mach-Zehnder interferometers. In inset thermal phase modulators control splitting ratio and differential output phase. Time ($\tau$) is defined from left to right; space ($i$) is defined from top to bottom.
		(b-c) Examples graphs implementable with the processor's Hamiltonian. Labeled dark circles represent waveguide positions (nodes).
		(b) Nearest-neighbor graph implemented in this work. The coupling between waveguides depends on whether the timestep is even or odd.
		(c) Binary tree graph allowed by the processor Hamiltonian.
		(d) Dynamic disorder is implemented by choosing $\{\phi_i\}$ such that there are no spatiotemporal correlations. Disorder strength is described via the $c_{d}$ parameter.
		(e) Static disorder is implemented by choosing $\{\phi_i\}$ to be constant in time, but uncorrelated in space.
		(f) Arrested transport of a quantum particles in strong, statically disordered system to external sites can be optimized by introducing dynamic disorder.
		}
		\label{fig:concept}
	}
	\end{figure*}
	\begin{figure*}[t]
	\includegraphics[width=7.2in]{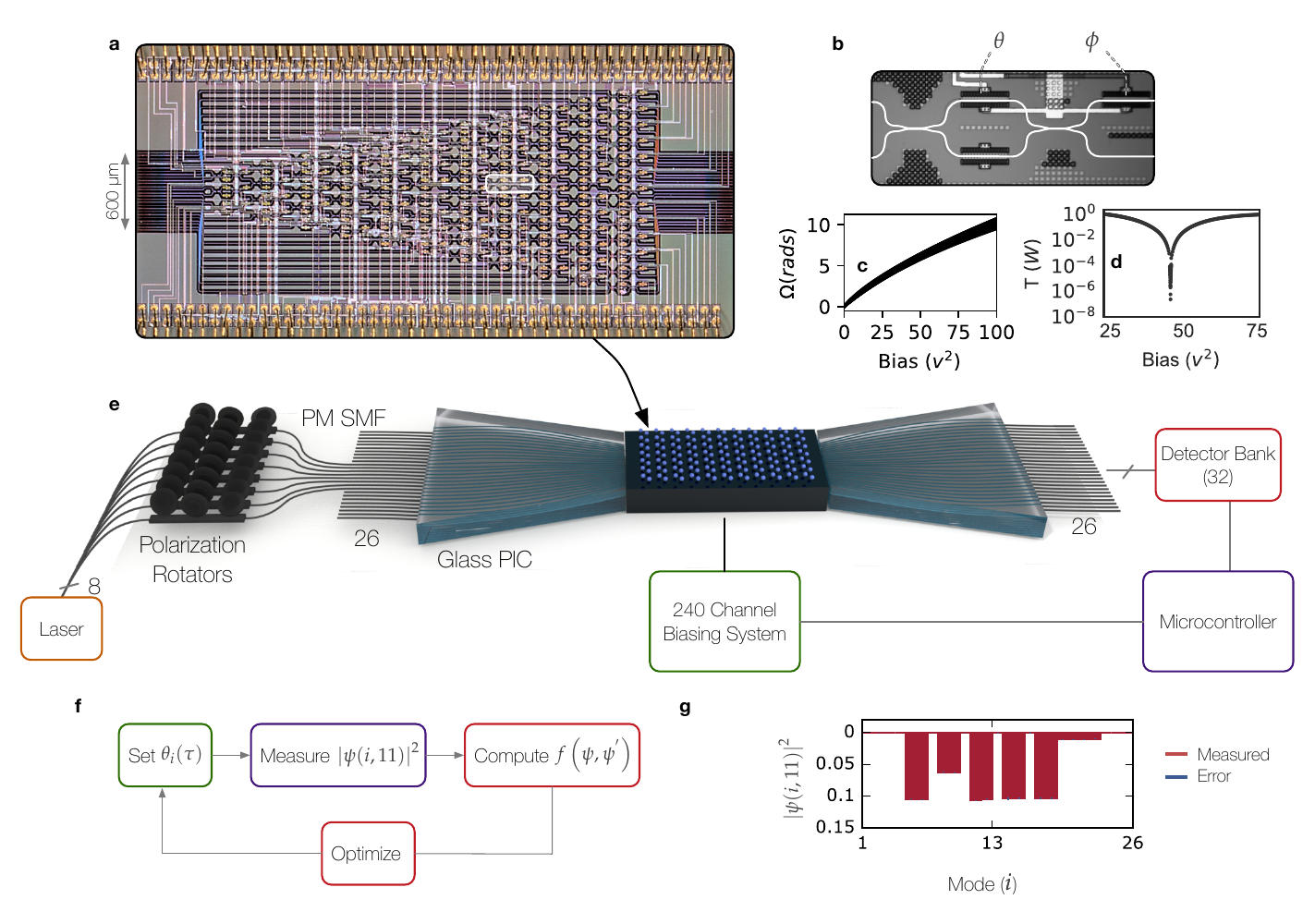}
	\centering
	\caption
	{
		\textsf
		{
		\textbf{Programmable nanophotonic processor.}
		(a) Processor composed of 88 Mach-Zehnder interferometers, 26 input modes, 26 output modes, and 176 phase shifters; gold wirebond filaments are visible.
		(b) Zoom in of the white inset in (a); shows zoom of individual interferometer and thermo-optic phase shifters ($\theta, \phi$).
		(c) Phase versus voltage curve for all internal and external phase shifters on the chip.
		(d) Transmission spectrum for an MZI with careful input polarization filtering. The extinction ratio is measured to be 0.9999998 $\pm$ 1.3$\cdot 10^{-8}$.
		(e) Setup used in this experiment. The processor accepts only transverse-electric polarized light, requiring a bank of polarization rotators to couple to the chip modes. Polarization rotator fibers are connected to the input glass photonic integrated circuit which serves as both a waveguide pitch reducer and as a spot-size converter. The output of the PNP is sent to an array of detectors and read out using a microcontroller. The processor is electronically programmed using a 240-channel biasing system and operated using a microcontroller.
		(f) Nonlinear optimization protocol used to generate the Massachusetts Institute of Technology logo shown in (g) across the 26 output modes of the processor $|\psi(i,11)|^2$.
		}
		\label{fig:micrograph}
	}
	\end{figure*}

	Quantum walks (QWs), the coherent analogue of classical random walks, have emerged as useful models for experimental simulations of quantum transport (QT) phenomena in physical systems. QT experiments have been implemented in various platforms including trapped ions\cite{schmitz2009quantum,zahringer2010realization}, ultra-cold atoms\cite{preiss2015strongly}, bulk optics\cite{PhysRevLett.104.153602,svozilik2012implementation,kitagawa2012observation,PhysRevLett.104.050502}, integrated photonics\cite{crespi2013anderson,schwartz2007transport,lahini2008anderson,Schreiber_2011,PhysRevLett.104.153602,Bromberg:2009ie,peruzzo2010quantum,PhysRevLett.108.010502,liu2013enhanced}, multi-mode fibers~\cite{defienne2016two}, and scattering media~\cite{wolterink2016programmable}. Integrated photonic implementations in silicon are particularly attractive for high interferometric visibilities, phase stability, integration with single-photon sources\cite{PhysRevX.4.041047,Collins_2013,silverstone2014chip} and detectors\cite{najafi2015chip}, and the promise of scaling to many active and reconfigurable components. The role of static and dynamic disorder in the transport of coherent particles has been of particular interest in the field of quantum simulation\cite{aspuruguzik2012photonic,huh2014boson}.

	Control over static (time-invariant) and dynamic (time-varying) disorder enables studies of fundamentally interesting and potentially useful quantum transport phenomena. In systems with strong dynamic disorder, a coherent particle evolving over $T$ time steps travels a distance proportional to $\sqrt{T}$; the coherent nature of the particle is effectively erased, resulting in classical, diffusive transport characteristics\cite{levi2012hyper,PhysRevE.79.050105}. In contrast, a coherent particle (or wave) traversing an ordered system travels a distance proportional to $T$ as a result of coherent interference between superposition amplitudes --- a regime known as ballistic transport. Perhaps most notably, a coherent particle propagating through a system with strong, static disorder becomes exponentially localized in space, inhibiting transport. This phenomenon, Anderson localization\cite{anderson1958absence}, has been observed in several systems, including optical media\cite{crespi2013anderson,lahini2008anderson,PhysRevLett.105.163905,Segev:2013ca,schwartz2007transport}. For systems in which transport has been arrested due to Anderson localization, it has recently been predicted that adding environmental noise (dynamic disorder), over a finite range of strengths, could result in enhanced transport. This effect, known as environment-assisted quantum transport (ENAQT), is believed to play a key role in the high transport efficiencies in photosynthetic complexes\cite{Rebentrost:2009hu,mohseni2008environment}. With such a variety of QT phenomena accessible, it is of fundamental interest to develop systems capable of fully probing the disorder space --- potentially revealing new QT phenomena in the single- and multi-particle regimes.

	In integrated photonic systems, static and dynamic disorder have been introduced by fabricating circuits with random parameter variations or post-processing\cite{crespi2013anderson,lahini2008anderson} --- rendering explorations of large parameter spaces in QT simulations possible, but cumbersome. Further, single instances of disorder are not suitable to characterize transport as they can produce a wide range of output distributions\cite{schwartz2007transport}; ensemble averages over many instances are necessary to accurately reproduce output statistics. While static and dynamic disorder have been studied separately\cite{schwartz2007transport,lahini2008anderson,Schreiber_2011,levi2012hyper,crespi2013anderson,PhysRevLett.104.153602}, transport in a system capable of implementing both simultaneously --- a requirement for the observation of ENAQT --- in all combinations and strengths, with low loss, and over a large number of instances has not been demonstrated. Developing systems capable of meeting these requirements, and the more stringent requirement of generating arbitrary single-particle unitary operations\cite{reck1994experimental}, has been the topic of several recent theoretical investigations\cite{miller2013selfconfiguring,mower2015high}.

	Here, we introduce a programmable nanophotonic processor (PNP), shown schematically in Fig.~\ref{fig:concept}(a), and leverage its programmability to explore the interaction between a particle undergoing quantum transport (on the graph depicted in Fig.~\ref{fig:concept}(b)) and two kinds of noise: static and dynamic disorder. Over a set of 64,400 experiments, we observe a number of hallmark quantum transport regimes including, for the first time, the signature of ENAQT in discrete-time systems and the ``quantum Goldilocks'' regime~\cite{lloyd2011quantum}. The generality of the PNP Hamiltonian unlocks further exploration into quantum transport on a large set of graphs including binary trees (shown in Fig.~\ref{fig:concept}(c)) and graphs that exhibit topological order~\cite{kitagawa2012observation}.

	Previous photonic integrated circuits for quantum information processing have been limited to 30 individually tunable elements and 15 interferometers\cite{carolan2015universal}. The PNP is composed of 176 individually tunable phase modulators and 88 interferometers spanning a chip area of 4.9~mm by 2.4~mm. This high component density is enabled by high silicon-to-silica index contrast (enabling the fabrication of waveguides with less than 15$\upmu$m bend radii (used here) and the large thermo-optic coefficient of silicon (enabling compact phase modulators\cite{harris2014efficient}).

	Recent work suggests that even sampling the output distributions from linear unitary processes, including QWs, with a relatively small numbers of photons ($n>30$ for a circuit with approximately 1000 modes) becomes intractable for classical computers\cite{aaronson2011computational,spring2013boson,broome2013photonic}. Further, QWs augmented with feed-forward control\cite{prevedel2007high,knill2001scheme,Kok_2007} have been proposed for universal quantum computation schemes\cite{UCQW2009}.
}
\section*{Programmable nanophotonic processor}
{
	The PNP, shown in Fig. \ref{fig:micrograph}(a), consists of a mesh of reconfigurable beamsplitters (RBSs, highlighted in white). Each RBS is composed of two 50\% directional couplers separated by an internal ($\theta$) and an external ($\phi$) thermo-optic phase shifter\cite{harris2014efficient}, as shown in Fig. \ref{fig:micrograph}(b). The RBS applies the rotation,
	$$
	\hat{U}_{RBS}=
	\left[\begin{matrix} e^{i \phi}\sin\theta		&		e^{i \phi}\cos\theta		\\		\cos\theta		 &		 -\sin\theta\end{matrix}\right]
	$$
	in the spatial-mode basis. Each RBS can implement any rotation in $SU(2)$ by using its internal and external phase shifter and the output phase shifter of the preceding RBS; for the input RBSs in the PNP, all of $SU(2)$ can be accessed by choosing an input phase.

	As shown in Fig.~\ref{fig:micrograph}(a), silicon waveguides in the PNP are inverse tapered to a mode-field diameter of 2$\upmu$m with a mode spacing of 25.4$\upmu$m. The 52 modes of the PNP are coupled to optical fibers using two laser written glass photonic integrated circuits as indicated in Fig.~\ref{fig:micrograph}(c) (see Supplement Section 1). Mode-field diameter mismatch between the glass chips and the PNP and fiber connectorization result in a transmission loss of 3.5~dB per facet. The total loss through the PNP, including both input and output coupling, is 8~dB. Accounting for coupling losses of 3.5~dB per facet the PNP transmission is 80\%; this matches the expected propagation loss for our silicon nanowire waveguides~\cite{baehr201225}.

	Each of the 176 phase shifters can impart more than 2$\pi$ radians phase shift as shown in Fig.~\ref{fig:micrograph}(d). The mean extinction ratio for the RBS unit cell is 0.9996 $\pm$ 0.0005 (limited by polarization drift, see Supplement Section 3 for unit cell measurement details). Fig.~\ref{fig:micrograph}(c) shows an extinction ratio of 0.9999998 $\pm$ 1.3$\cdot 10^{-8}$ (approximately 66.3~dB) for a single MZI measured with the test setup shown in Supplement Section 3. Nearest-neighbor thermal cross-talk between adjacent phase shifters was measured to be 1\%. We correct for this by performing a linear matrix inversion for each program. The thermo-optic modulators have a 3~dB bandwidth of 130.0 $\pm$ 5.59 kHz\cite{harris2014efficient}, permitting up to 10$^5$ full PNP reconfigurations per second.

	The PNP was fabricated in a complimentary metal-oxide semiconductor (CMOS) -compatible, silicon photonics process (see Methods for more detail). After fabrication, we calibrate the PNP as described in Supplement Section 4. To program all 176 phase shifters, we developed a 240-channel, 16-bit precision electronic biasing system. To measure the PNP output mode intensities, we developed a 32-channel detector array system with 18-bit readout precision. See Supplemental Section 2 for experimental setup details. As a demonstration of the programmability of the PNP, we used a nonlinear optimization algorithm to learn the phase settings required to generate the Massachusetts Instutitue of Technology (MIT) logo across the output modes of the PNP ($|\psi(i,11)|^2$) as shown in Fig.~\ref{fig:micrograph}(f,g)---for details, see Supplement Section 7.
}
\section*{Experiment}
{
	The PNP is described by the following interaction Hamiltonian:
	$$
	\hat{\mathcal{H}}_{int} \propto
	\sum_{i} \frac{i\theta_{i}(\tau)}{2}
	\left( \hat{a}_i \hat{a}_{i+1}^{\dagger} -
	\hat{a}_{i+1} \hat{a}_{i}^{\dagger} \right) +
	i \phi_{i}(\tau) \hat{a}_i \hat{a}_{i}^{\dagger}
	$$
	where $\hat{a}$ is the mode annihilation operator, $\tau$ from 1 to 11 is the MZI column number (or time step), and $i$ between 1 and 26 is the waveguide number as shown in Fig.~\ref{fig:concept}(a). This Hamiltonian is more general than those demonstrated to this point, enabling simulations of transport on a range of graphs including binary trees such as the one in Fig.~\ref{fig:concept}(c) originally considered by Rebentrost et al.~\cite{Rebentrost:2009hu}. Here, we will explore the case $\theta_i(\tau) = \pi/2$---equal coupling between sites. The graph implemented by these settings is shown in Fig.~\ref{fig:concept}(b). To realize transport distributions that are symmetric about the initial site at $i=14$, we program $\phi_{14}(1)$ to $\pi/2$~\cite{kempe2003quantum}. To pause the evolution of the particle at some time-step $\tau_0$, we can set $\theta_{i}(\tau > \tau_0) = \pi$.

	As illustrated in Fig.~\ref{fig:concept}(d), dynamic disorder is added to the system by sampling $\phi_i(\tau)$ (for all $i$ and $\tau$) from the uniform distribution with values between $\pi$ and $-\pi$. This disorder results in a particle phase evolution that is uncorrelated in both position and time. Static disorder, illustrated in Fig.~\ref{fig:concept}(e), is added to the system by sampling $\phi_i$ (for all $i$) from the uniform distribution between $\pi$ and $-\pi$. This results in a particle phase evolution that is uncorrelated in space, but constant in time. We use distribution weighting parameters $c_{d}$ and $c_{s}$ between 0 and 1 to control dynamic and static disorder strength, respectively. We define \textit{disorder coordinates} as $(c_{d}, c_{s})$.

	\begin{figure*}[t]
	\includegraphics[width=7in]{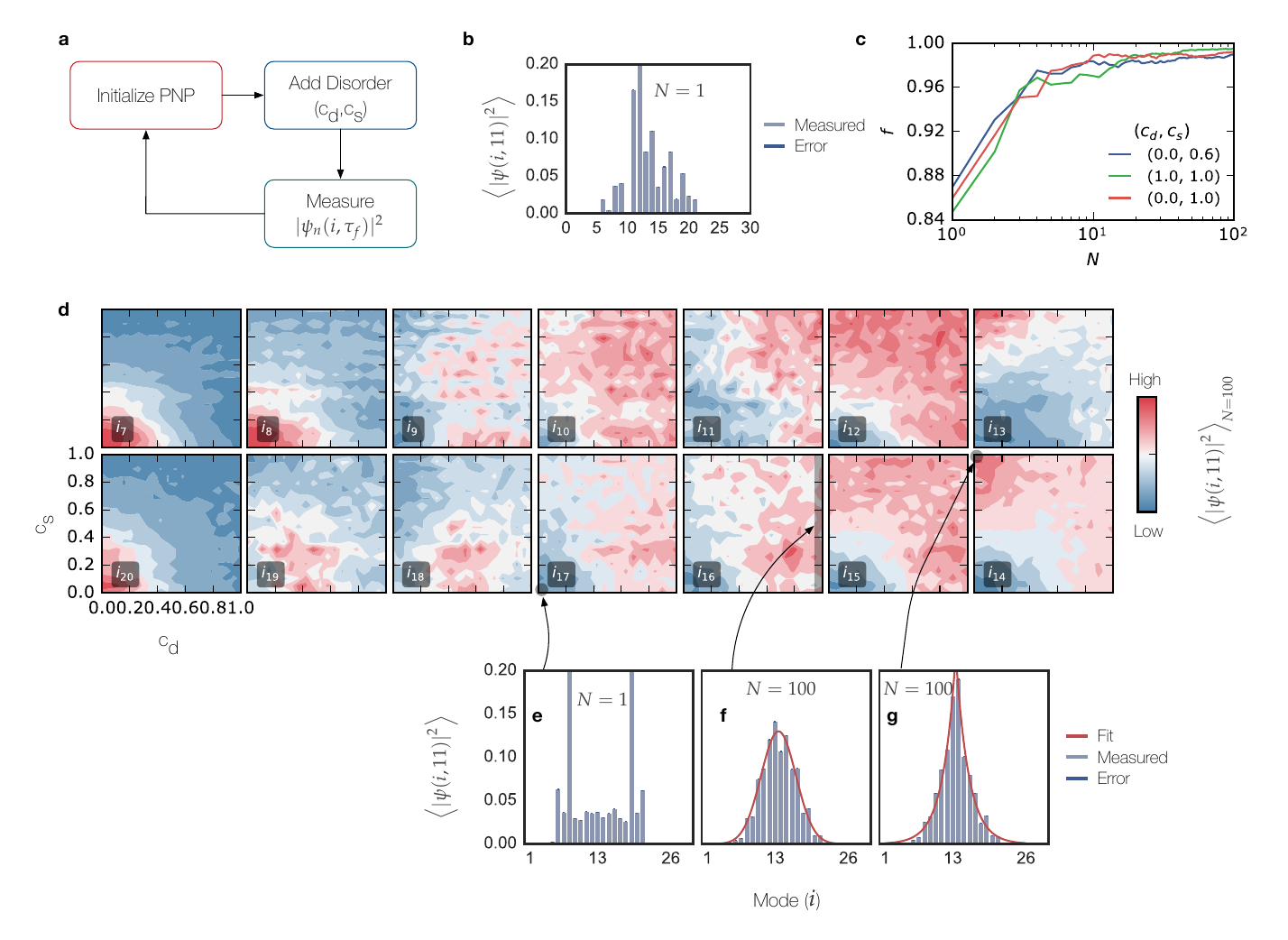}
	\caption
	{
		\textsf
		{
		\textbf{Convergence and the full noisy transport space.}
		(a) Programming routine for transport experiments. PNP is initialized, environmental noise is introduced, and the output distribution is measured. This process is repeated $N$ times to develop robust statistics.
		(b) A measurement of a single instance of disorder coordinate (0, 1). With one instance, it is difficult to determine what disorder coordinate could have generated this distribution.
		(c) Convergence of the mean output distribution with iteration number. $N \geq 10$ to achieve a fidelity $f$ exceeding 98\%.
		(d) $\mean{\psi}{i}{11}$ for 400 combinations of $(c_{d}, c_{s})$ with both disorder coordinates varying on $[0,1]$. Modes labeled $i=7\rightarrow 20$ are shown; $i<7$ and $i>20$ are relatively constant for all levels of disorder, as predicted via simulation. The mean fidelity for all modes and all disorder coordinates is 99.8 $\pm$ 0.036 \%. Particle-like, incoherent transport occurs on the right edge of these plots. Coherent, ballistic transport occurs in the bottom left corner.
		(e-g) Measurements of $\mean{\psi}{i}{11}_{N=100}$ at disorder coordinates (0,0), (1, 1), and (0,1). Fits to the Laplace and Gaussian distributions shown in red in (f) and (g), respectively.
		}
		\label{fig:space}
	}
	\end{figure*}

	\begin{figure*}[t]
	\centering
	\includegraphics[width=7in]{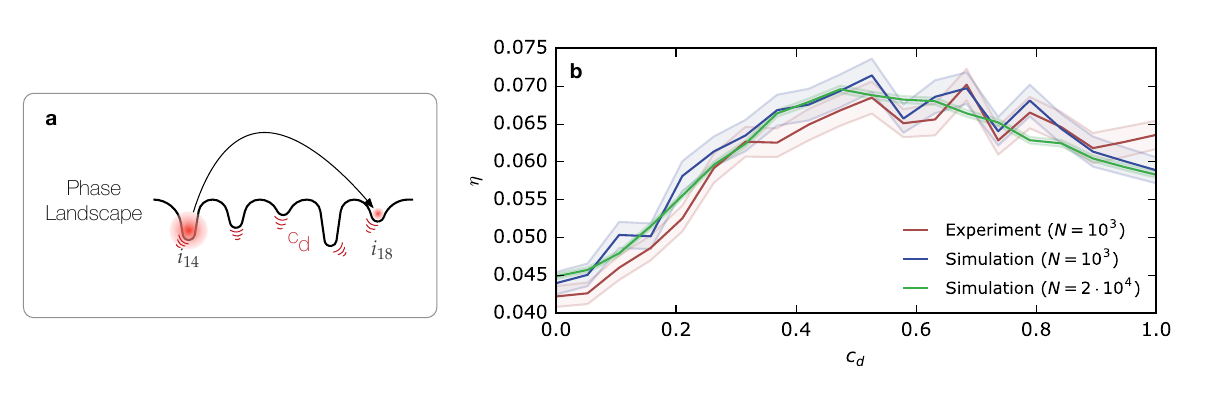}
	\caption
	{
	\textsf
	{
	\textbf{Environment-assisted quantum transport and the Goldilocks regime.}
	(a) Conceptual drawing of the phase landscape for a strong, statically disordered system where particle is localized to initial site at $i_{14}$. By adding dynamic disorder (shown as red vibrations), it is possible to optimize transport of the particle to further-away sites.
	(b) Slice through $c_s=0.6$ and all $c_d$ corresponding to a randomized phase span of $1.2\pi$. Y-axis represents quantum transport efficiency $\eta$, x-axis is dynamic disorder strength. Red and blue lines are experimental data and simulation data for the phase configuration used in the experiment, respectively. Green line represents simulation with $N=2\cdot 10^4$---the asymptotic distribution. These data show environment-assisted quantum transport in a discrete-time system. Standard errors are plotted as a transparent band around the mean.
	}
	\label{fig:contour}
	}
	\end{figure*}

	\subsection*{Mapping the environmental interaction space}

	It is important to note that these photonic simulations of environmental noise produce deterministic output distributions. By generating many systems described by the same disorder coordinate, general transport behavior for a system described by this coordinate can be simulated. As shown in Fig.~\ref{fig:space}(a), we perform a single QT experiment by (1) initializing the PNP to $\theta_{i}(\tau)=\pi/2$ and $\phi_{14}(1)=\pi/2$, (2) programming a single instance ($n$) of a given disorder coordinate, and (3) making a measurement of the output distribution $\intensity{n}{i}{\tau}$ and then normalizing this distribution to sum to one (see Methods). To illustrate the need for averaging over many instances, we measured $\intensity{n}{i}{\tau}$ for one instance of a disordered system described by (0,1), as shown in Fig.~\ref{fig:space}(b). With only a single instance, it is unclear what disorder coordinate could have generated this distribution or even whether the particle undergoing transport is coherent.

	To benchmark the number of instances required for space of the mean distribution $\mean{\psi}{i}{11}= \frac{1}{N} \sum_{n}\intensity{n}{i}{11}$ over a number of instances ($N$), we program the PNP to implement systems with disorder coordinates (0, 0.6), (1.0, 1.0), and (0, 1.0) and measure output distributions for $N=10^2$ instances of each coordinate (see Supplement Section 5 for measurements at each $\tau$). Next, we use a computer to simulate the same coordinates with $N=2\cdot 10^4$ instances---we call this the `asymptotic distribution'. For each $n \leq N$, we compare the fidelity (see methods for fidelity metric) of $\mean{\psi}{i}{11}$ up to the $n$th measurement against the asymptotic distribution. The data is plotted in Fig.~\ref{fig:space}(c). $N \geq 10$ instances are required to converge to the asymptotic distribution with a fidelity exceeding 98\% --- we use $N\geq 10^2$ for all following experiments.

	\begin{table}[h]
	\begin{tabular}{l l l}
	 & Fidelity & Uncertainty \\
	 \hline
	Fig~\ref{fig:space}(b)	& 0.9990	& $\pm$ 0.0024\\
	Fig~\ref{fig:space}(d)	& 0.9976	& $\pm$ 0.0004\\
	Fig~\ref{fig:space}(e)	& 0.9988	& $\pm$ 0.0030\\
	Fig~\ref{fig:space}(f)	& 0.9987	& $\pm$ 0.0035\\
	Fig~\ref{fig:space}(g)	& 0.9989	& $\pm$ 0.0035\\
	\end{tabular}
	\caption[test]{Fidelities for experimental data shown in Fig.~\ref{fig:space}.}
	\label{table:fids}
	\end{table}

	To probe whether ENAQT can be observed in discrete-time systems, we measure 400 disorder coordinates with $c_{d}$ and $c_{s}$ each spanning all strengths from 0 to 1 in increments of 0.05 and $N=100$---totaling $4\cdot 10^4$ experiments. The ensemble average probability distributions at the output of the PNP for modes 7-20 are plotted in Fig.~\ref{fig:space}(d); modes 1-6 and 21-26 are not plotted since transport to these extremal modes is nearly uniform for all disorder levels as predicted by simulation. The measured data in Fig.~\ref{fig:space} is in excellent agreement with simulation as shown in Table~\ref{table:fids}. To the best of our knowledge, these are the highest fidelities reported for quantum transport experiments. This highlights one of the key benefits of the PNP: it can be calibrated post fabrication to implement high fidelity transformations without physical modification.

	To gain intuition for the disorder space in Fig.~\ref{fig:space}(d), we have plotted $\mean{\psi}{i}{11}$ for disorder coordinates (0, 0), (1, 1) and (0, 1) in Fig.~\ref{fig:space}(e-g), respectively. The lower left coordinates in the Fig.~\ref{fig:space}(d) correspond to ballistic transport (Fig.~\ref{fig:space}(e)), the right edge corresponds to diffusive, incoherent transport (Fig.~\ref{fig:space}(f)), and the upper left coordinates correspond to Anderson localization (Fig.~\ref{fig:space}(g).

	\subsection*{Environment-Assisted Quantum Transport}
	{
		Of particular interest in these data is the behavior of mode 18 in Fig~\ref{fig:space}(d). We examine a cross-section through $c_{d}$ between 0 and 1, with $c_{s}=0.6$--corresponding to a strong, statically disordered system with $\phi_i$ sampled uniformly between $\pm 0.6 \pi$. In this case, a coherent particle localized at the starting point $i_{14}$ is able to escape its initial position through the introduction of dynamic disorder. This process is depicted schematically in Fig.~\ref{fig:contour}(a). The cut through this transport space is plotted in Fig.~\ref{fig:contour}(b); the experimental and simulated data are in close agreement with a fidelity of 0.9998 $\pm$ 0.0157. Intuitively, one may expect that introducing further dynamic disorder would enhance transport to site $i_{18}$. The data indicates that this is not the case. Instead, there is a "quantum Goldilocks"~\cite{lloyd2011quantum} regime in which adding additional dynamic disorder inhibits transport. The maximum efficiency improvement gained through ENAQT in this system after 11 time steps ($\Delta\eta = \left(\eta_{max} - \eta_{min}\right) / \eta_{max}$) is measured to be 42\%.
	}
}

\section*{Discussion}
{
	ENAQT occurs in (statically) disordered systems in which transport has been arrested due to Anderson localization; through interactions with a fluctuating environment (simulated by dynamic disorder), the transport efficiency of the coherent particle may be enhanced. Rebentrost et al.\cite{Rebentrost:2009hu} predicted the existence of such an effect in systems that evolve in continuous-time. We have shown the first evidence for ENAQT on discrete-time graphs as well as a "quantum Goldilocks"~\cite{lloyd2011quantum} regime in which there is an optimal transport efficiency for all levels of environmental noise. We note that the general Hamiltonian of this processor enables a range of quantum transport experiments on a variety of graphs.

	For multi-photon experiments, loss is important. Coupling loss in our setup could be reduced to less than 5\% using grating couplers~\cite{notaros2016ultra}. Using improved fabrication, waveguide losses can also be lowered considerably; values as low as 0.3~dB/cm have been reported~\cite{Cardenas:09}. With these improvements, the transmission loss of a photonic system of the complexity used here could be reduced to 9\% using current technology. To the best of our knowledge, our MZI visibility (over 66~dB) is the highest reported in literature. This indicates that developing RBS unit cells composed of multiple MZIs (to combat fabrication defects) may be unnecessary~\cite{wilkes201660}.

	For these experiments, we carefully model and calibrate the PNP to maximize fidelities for single-particle experiments---enabling future high-fidelity, multi-photon quantum transport experiments (see Supplement Section 6 for two-photon quantum interference experiments using the PNP). Transport simulations exceeding several tens of photons\cite{Yao_2012} become intractable\cite{aaronson2011computational} for classical computers. While we have demonstrated 1D quantum transport here, it is possible to simulate unitary processes in higher dimensions (e.g. 2D, 3D) by embedding circuits~\cite{carolan2015universal} capable of implementing arbitrary unitary operators. Entangled photon sources\cite{harris2014efficient,Collins_2013,silverstone2014qubit} and single photon detectors\cite{najafi2015chip,gerrits2011chip} have recently been integrated into the silicon photonics platform, providing a path towards high photon number experiments. In addition, recently demonstrated low-latency superconducting logic devices\cite{McCaughan_2014} may provide fast on-chip feed-forward operations on quantum optical states. Together with programmable nanophotonic processors, these chip-integrated technologies provide a promising platform for future quantum simulation and computing tasks.
}

\section*{Methods}
{
	\small
	\textbf{Fabrication of the photonic circuit.} We fabricated the PNP on a silicon-on-insulator wafer with 2~$\upmu$m of buried SiO$_2$ in the Optoelectronic Systems Integration in Silicon foundry. Device regions are defined using partial- and full-etches in the silicon device layer resulting in 90~nm and 220~nm thick structures, respectively. A silica cladding 2$\upmu$m thick is thermally grown on top of the device layer to achieve transverse symmetry for the optical modes. Input and output optical coupling is achieved with an inverse taper from 500~nm wide to 200~nm wide over a distance of 300$\upmu$m. Light is guided in 500~nm wide by 220~nm thick silicon ridge waveguides with a single transverse-electric mode (TE$_0$) near 1550 nm, an effective index of 2.57. Thermo-optic phase shifters\cite{harris2014efficient} are defined by full- and partial-etches and two Boron implants with concentrations 7$\cdot$10$^{17}$ cm$^{-3}$ and 1.7$\cdot$10$^{20}$ cm$^{-3}$. Due to high active element densities, two aluminum layers and two via layers are required to simplify the phase shifter electrical signal routing.

	\small
	\textbf{Measurement description.} Continuous-wave laser light (100$\upmu$W at 1570~nm wavelength) is launched at input 12 and intensity at outputs 1-26 are measured using an array of calibrated photodiodes. Loss and coupling are independently measured for each input and output channel to enable accurate estimation of the resultant output intensity distributions. The measured intensity distributions are normalized to give the probability distributions presented.

	\small
	\textbf{Fidelity calculations.} To evaluate the closeness between 1D distributions, we use the metric $\sum_i \sqrt{p_i q_i}$ where $p$ and $q$ are probability-normalized distributions (i.e. $\sum_i p_i=1$). For 2-D distributions, we use the metric $\sum_{ij} \sqrt{P_{ij} Q_{ij}}$ where the distributions are normalized to unity as $\sum_{ij} P_{ij} = 1$.

	\small
	\textbf{Data availability.} The data that support the plots within this paper and other findings of this study are available from the corresponding author upon reasonable request.
}

\section*{Author Contributions}
{
	\small
	N.H. designed the photonic integrated circuit, and experimental setup and performed the experiment. N.H. laid out the design mask with assistance from G.S. on metal routing. N.H., G.S., Y.L. S.L. and D.E. conceived the experiment. T.B.J. and M.H. fabricated the system. All authors contributed to writing the paper.
}

\bibliography{citations}

\section*{Acknowledgments}
{
	\small
	N. H. acknowledges support from the National Science Foundation Graduate Research Fellowship grant (1122374). G.S. acknowledges support from the Department of Defense National Science and Engineering Graduate Fellowship. D.E. acknowledges support from the Sloan Research Fellowship in Physics. Y.L. acknowledges support from the Pappalardo Fellowship in Physics. This work was supported in part by the Air Force Office of Scientific Research (AFOSR) Multidisciplinary University Research Initiative (FA9550-14-1-0052) and the Air Force Research Laboratory program (FA8750-14-2-0120). M.H. acknowledges support from AFOSR small business technology transfer program, (FA9550-12-C-0079, FA9550-12-C-0038) and Gernot Pomrenke, of AFOSR, for his support of the optoelectronic systems integration in silicon (OpSIS) effort, though both a Presidential Early Career Award in Science and Engineering award (FA9550-13-1-0027) and funding for OpSIS (FA9550-10-1-0439). The authors thank Christophe Galland for his helpful discussions of the result.
}
\end{document}